# Science, Technology and Non Equilibrium Statistical Mechanics

Clóves Gonçalves Rodrigues[1, *], Roberto Luzzi[2]

[1]School of Exact Sciences and Computing, Pontifical Catholic University of Goiás, Goiânia, Brazil
[2]Institute of Physics "Gleb Wataghin", University of Campinas, Campinas, Brazil

**Abstract**

The binomial *Science & Technology* (S & T) is inseparable. In this contribution are presented some general considerations on the question of the aspect and interrelationships of the Science, Technology, Government and Society, and the role of *Non Equilibrium Statistical Mechanics* that are present in the nowadays highly sophisticated technologies and industrial processes that contribute to the wealth and well being of world society. The evolution of the Non Equilibrium Statistical Mechanics is briefly described. In present days remarkable development of all the modern technology, essential for the welfare and progress of the global society, imposes an immense stress on basic Physics, and consequently on Non Equilibrium Statistical Mechanics, in situations like, for example: fluids with complex structures, electronics and photonics involving systems out of equilibrium, nano-technologies, low-dimensional systems, non-linear and ultrafast processes in semiconductors devices, and soft matter. It is shown that the Non Equilibrium Statistical Mechanics can deal, within a certain degree of triumph, with some these situations, namely: 1. nonconventional thermo-hydrodynamics; 2. ultrafast relaxation processes in semiconductors devices; 3. nonequilibrium Bose-Einstein-like condensations and coherent states; 4. low-dimensional semiconductors; 5. thermo-statistics of complex structured systems; 6. nonlinear higher-order thermo-hydrodynamics; and 7. nonlinear transport in semiconductors devices. These areas are of particular interest, at the scientific, technological and at the production line, and therefore of relevance for government and society, successfully analyzed in terms of the formalism of Non Equilibrium Statistical Mechanics.

## 1. Introduction

The binomial *Science & Technology* (S&T) is inseparable. It is said, quite rightly, that Science is the mother of technologies, which, in a symbiotic process, are involved in a positive feedback mechanism. For example, in this context, Thermodynamics, a scientific discipline which deals with the connection between heat and mechanical work, can be considered an offshoot of the Industrial Revolution - initiated in the United Kingdom in the XVIII century for in continuation to get extended through most of the world - in the sense that technology propitiates or, better to say, forces the improvement of Science. It is also worth emphasizing that the binomial S&T is the locomotive of the world economy and therefore of the wealth of nations and the well being of world society. The economic growth of nations and therefore, as said, the corresponding well being of their societies, is expected to be enhanced when S&T is accompanied with what is dubbed as the "mantra" of *Innovation*. It has been stated that innovation should incorporate not just technological innovation, but social innovation and also nurturing innovative people, in what society has to become more conducive to innovation and provide opportunities for risk-taking, adventurous people. Also, that even though investments in science and technology provide the seeds for economic value, in this globalization age scientists have to compete and deliver the seeds of scientific discovery to the market-place: That requires social encouragement of entrepreneurial activities, radically increasing the active participation of Universities [1]. This has to do with "Globalization", which is considered that has introduced both uncertainties and opportunities world wide. It is introduced an "innovation ecology" consisting of interrelated institutions, laws, regulations, and policies providing an innovation infrastructure that entails education, research, tax policy, intellectual property protection, among others [2].

A fundamental pillar for *Science, Technology and Innovation* consists in the creation and diffusion of knowledge, and C. H. Brito Cruz [3] has noticed that knowledge is more and more becoming a main commodity for the generation of wealth and social well-being. The capacity of a nation to generate knowledge and transform it in prosperity and social development depends on the action of some institutional agents which originate and apply knowledge. The main agents are business, universities and government.

The role of governments was deeply analyzed by a panel of renowned people in the Carnegie Commission on Science, Technology, and Government (New York, USA), producing the Report Science, Technology, and Government for a Changing World (1993). In the Preamble, the distinguished professor Joshua Lederberg (decease in 2008) wrote that:

"Government is the complex of institutions, laws, customs, and personalities through which a political unit exercises power and serves its constituencies. Science is the search for novel and significant truths about the natural world. These truths are usually validated by the prediction of natural phenomena and the outcome of critical experiments. Technology is the instrumental use of scientific knowledge to provide, for example, goods and services necessary for human sustenance and comfort and to support other, sometimes contradictory aims of the political authority. Scientific expertise and technology have always been valued by government. Weapons and medicines, maps and microprocessors: the products of science are indispensable to successful government. So, increasingly, is scientific thinking. Where but to science can society turn for objective analysis of technical affairs? The science mind bring much to the political processes. But science and politics are a hard match. Truth is the imperative of science; it is not always the first goal of political affairs. Science can be, often should be, a nuisance to the established order, much as technology often bolsters it. Moreover, many scientists, lacking the policy skills needed to relate their expertise to social action, are uncomfortable dealing with the political machinery. A vital responsibility of the expert advisor is to clarify technical issues so that the essential policy questions become accessible to the judgment of the community at large. Yet expertise also has distortions, arising from conflicts of interest, differing levels of competence, peculiarly posed questions, and cultural biases. The discipline of the peer group is the main source of the authenticity of the scientific community. Science, in fact, cannot exist without a community of scientists, a forum for organized, relentless skepticism of novel claims. Science kept in confidence and inaccessible to colleagues' criticism is no longer authentic. The public rendering of advice and defense of conclusions is of the utmost importance. Nevertheless, advice within the political system must often be confidential. Herein lies another structural contradiction and challenge to the design of organization and decision making. We must thus establish institutions and processes that enable scientists both to be credible within polities and to remain worthy of the continuing confidence of the larger society. To achieve this dual goal the first social responsibility of the scientist remains the integrity of science itself. Scientists fear that a greater influence on policy will evoke more explicit political control of science. A healthy balance is in the interests of both science and government."

In that report it has also been noticed that more than half a century ago it was provided by Vannevar Bush, science advisor to President Franklin Roosevelt, a so called visionary report, titled "*Science - The Endless Frontier*", on the future of science and technology. At the time it was published, the Second World War – the driving force behind many scientific and engineering accomplishments – has just ended and the United States faced fundamental questions about the interactions of universities, industry, and government in furthering science and technology. In Bush's words, "The government should accept new responsibilities for promoting the flow of new scientific knowledge and the development of specific talent". Science, Bush argued, should serve society, and in turn, society should provide the financial support to assure the advancement of science, particularly basic research. It may be added that the so-called Cold War was also a phenomenal driving force for an enormous development of science and technology. Suffice to notice, for example, the development of space satellites and the voyage to the Moon, and, of course, the Internet Network. On the role of governments John Marburger, Science and Technology Advisor to the USA presidency during the administration of George W. Bush, has noticed that science policy depends on the state of science itself, which evolves in response to new instrumentation, theoretical methods, and analytical tools, including digital computing. The growth of science and the course of science policy are undeniable progressive. Thus science policy necessarily depends to a great extent on the state of science itself, and not only on social conditions and the willingness of governments to fund research. The accumulation of knowledge (as noticed above [3]) guarantees that science's future will differ from its past. Fields mature, saturate, and merge as the frontiers of discovery advance. Attitudes toward science and toward physics in particular, have been shaped by the immense fertile period when quantum mechanics was first seriously exploited in the decades after World War II. But that period is now behind us and attitudes are necessarily changing [4].

It is certainly a truism to say that, at present, the society is witnessing a tremendous development of technologies and, of course, the corresponding production of goods and materials that are fantastically benefiting large portions of people in the world.

Rodger Doyle [5] has noticed that these recent inventions are sometimes hailed as a "Third Industrial Revolution". The "First Industrial Revolution" (~1770's to ~1860's) saw the development of the steam engine, steamboat, locomotive, telegraph, cotton gin, steel plow among others. The "Second Industrial Revolution" (~1870's to ~1910's) witnessed the invention of the telephone, internal combustion engine, the electric light bulb, germ theory of disease, linotype, motion picture, radio, air conditioning, airplane, indoor-flush toilet among others. R. Doyle advance the conclusion that the first and second Industrial Revolutions led to fundamental changes in human affairs, which have been not rivaled by this so-called "Third Industrial Revolution", which then is "Not So Revolutionary" (see also Ref. [6]).

Leaving this point aside, Mary L. Good, USA Undersecretary of Commerce in the administration of Bill Clinton, noticed that nowadays much of the world is waking up to the economic promise of technology in the present-day globalization scenery: Technology is directly linked to the economic growth of nations. The globalization of technology poses difficult challenges for policymakers, not just in technology policy, but in others such as trade and regulatory policy [7].

On this, the 1992-Report of the Carnegie Commission *Enabling the Future: Linking Science and Technology to Societal Goals* [8], begins the Executive Summary with a citation of Antoine de Saint-Exupèry (in *The Wisdom of the Sands*): "As for the Future, your task is not to foresee, but to enable it".

It is also stated that: "Basic scientific research is a voyage of discovery, sometimes reaching the expected objective, but often revealing unanticipated new information. Some might say that setting long-range goals may harm basic researchers by overcentralizing and removing flexibility from the system. Long-range S&T goal setting certainly should not hamper, but rather encourage, this freedom to discover. Knowledge resulting from basic research must be exploited to improve the efficiency and effectiveness with which applied research and technological development are directed to societal goals". On the latter, in page 24 of the Report are listed examples of major societal goals to which S&T contribute: They are subsumed in 4 general groups, namely,

1. Quality of Life, Health, Human Development, and Knowledge;

2. A Resilient, Sustainable, and Competing Economy;

3. Environmental Quality and Sustainable Use of Natural Resources;

4. Personal, National, and International Security.

The Commissioners also noticed that: Policy questions will not be resolved by citizens, scientists, business executives, or government officials working alone; addressing them effectively will require the coordinated effort of all sectors of society. As President John Kennedy said: "*Scientists alone can established the objectives of their research, but society, in extending support for science, must take account of its own needs*".

C. H. Brito Cruz [9] called the attention to the question of S&T in the Brazilian State of São Paulo, Brazil, enormously pushed forward by the creation of the São Paulo State Research Foundation (FAPESP). In that article it is cited the beginning of the document "Science and Research" prepared in 1947 for the advice of the members of the Assembly preparing the Constitution of the State; in a tentative translation it reads: "*Science assumes a function more and more preponderating in the destiny of Humankind* [...] *In peace, it is Science that provides orientation to economy and industry and promotes the greatness and well-being of nations.*"

Brito Cruz notes that the actuality of the argument is flagrant; and its efficacy was modular: from it was born the idea to create FAPESP. It is worth noticing that at the Brazilian federal level, in the decade of the 1950's and beginning of the 1960's, the National Research Council (CNPq), under recommendations of, in particular, Profs. Mario Schenberg and Jayme Tiommo, takes initiatives as, for example, the development of the Solid State Physics Laboratories of the São Paulo State University (USP). It can also be mentioned the important development in further introducing laboratories involving condensed matter physics, by Sergio and Yvone Mascarenhas in USP-SC. A decisive improvement followed after the second half of the 1960's with, within the context of the so-called "pluriannual plans" – through the action in the Ministry of Planning of ministers Roberto Campos, Helio Beltrão and João Paulo dos Reis Veloso – the government decision to create Graduate Centers of Excellence in Science and Technology with the help of the financing agency Finep. Presently, a Brazilian government pluriannual plan for improvement of Science, Technology & Innovation can be consulted at www.mct.gov.br.

Nowadays, let it be the "Third Industrial Revolution" or a large evolution of the "Second Industrial Revolution", the remarkable development of the "advanced modern Technologies" and intense "research and development", ask for Physics to bring to the forefront the *Physics of Systems out of Equilibrium* [10-12] and the *Physics of Non-Linear Processes* [13]. Currently it is necessary to also consider, among other disciplines, the mesoscopic physics [14], the physics of fractal structures, soft matter [15-17], ultrafast processes [18, 19], synergetic and self-organization associated to complex systems [20-22].

In present days remarkable development of all the modern technology, essential for the welfare and progress of the global society, imposes an immense stress on basic Physics, and consequently on Thermo-Statistics, in situations like, for example: fluids with complex structures, in electronics and photonics involving systems out of equilibrium, nano-technologies, low-dimensional systems, non-linear and ultrafast processes in semiconductors devices, and soft matter. All these topics are important for technological improvement in industries like, for example: in medical engineering, petroleum, food, polymers, cosmetics, electronics and optoelectronics. It is then required to introduce a thermo-hydrodynamics going well beyond the classical thermo-hydrodynamics. In the situations above mentioned there often appear difficulties of description, which impair the proper application of the conventional ensemble approach used in general, for example, the Boltzmann-Gibbs statistics. One way, to partially overcome such difficulties, is to use a non-conventional approaches.

Nowadays, the commercial interests of the technology industry claims for miniaturization of electronic devices, and then this rises the question if the understanding of the physics of electronic devices and their functioning can be extrapolated to the ultra-short time and ultra-small space [18, 19].

Statistical Mechanics can provide the basic scientific foundations for answering some of the questions above, more precisely, "Nonequilibrium Statistical Mechanics" whose evolution is described in the next Section.

## 2. The Evolution of the Non Equilibrium Statistical Mechanics

Oliver Penrose noted that statistical mechanics has conceptual problems with difficult questions to give a good answer [23]. However, the Gibb's ensemble algorithm provides a precise description for large physical systems in equilibrium. However, in the case of systems out of equilibrium, the Gibb's ensemble algorithm is not accurate.

The Statistical Mechanics is a theoretical construction that superseded the kinetic theory of the century XIX [24]. The theory of Gibbs looks for the physical and the conceptual aspects and has a fundamental foundation, with microscopic basis of phenomenological thermodynamics. But, the theory of Gibbs went beyond that, trying to describe all the macroscopic physical properties of the systems from a microscopic level by also providing basic foundations to response function theory. The construction of the Gibbs method is described in the textbooks in scheme orthodox: deterministic and reversible mechanics associated with *ad hoc* hypotheses.

The Probability Theory appears to be an *indispensable necessity* to describe phenomena at the macroscopic level. R. Feynman noted that, it is not our ignorance of the *internal mechanisms* that makes nature has probabilistic character: it seems to be intrinsic [25]. In this sense, J. Bronowski noticed:

"The future does not already exist: it can only be predicted" [26].

Several scientists have emphasized that the concept of probability is fundamental to the development of the science, including the scientific study of dynamic systems, let it be physical, biological, archeological, chemical, economic, ecological, social, historical, etc. According to J. Bronowski: "... *The statistical concept of chance may come as dramatically to unify the scattered pieces of science future...*" [26].

The main objective of Statistical Mechanics of many-body systems out of equilibrium is to determine their dynamical evolution and the thermodynamic properties of their macroscopic observables, in terms of the dynamical laws which govern the motion of their constitutive elements. This implies in:

1. to build an irreversible thermodynamics;

2. to build a response function theory and a generalized nonlinear quantum kinetic theory, which are fundamental to connect theory with experiment [27].

Oliver Penrose noted that statistical mechanics has conceptual problems with difficult questions to give a good answer [23], for example: "How can we justify the standard ensembles used in equilibrium theory? What is the physical significance of a Gibbs' ensemble? What are the correct ensembles for systems out of equilibrium? How can we reconcile the irreversibility of macroscopic behavior with the reversibility of microscopic mechanics?"

R. Kubo announced that: "the statistical mechanics of nonlinear systems is in its infancy, and further progress can only be hoped by closed cooperation with experiment" [28]. Moreover, the study of the systems out of equilibrium is more difficult than those in equilibrium. This difficulty is mainly due that a more detailed analysis is necessary to determine the temporal dependence of macroscopic properties, and also to calculate transport coefficients which are time- and space-dependent. According to R. Zwanzig the objectives of nonequilibrium statistical mechanics are:

1. to obtain transport equations and to understand their structure;

2. to obtain the temporal evolution and instantaneous values of the macroscopic quantities of the system;

3. to obtain the properties of the steady state, and

4. to understand how the approach to equilibrium occurs in natural systems [11].

Furthermore, Robert Zwanzig pointed out that for the purpose to face these questions there exist many approaches. These approaches can be classified as:

1. generalizations of Gibbs' ensemble,

2. expansions from an initial equilibrium ensemble,

3. techniques based on the generalization of the "theory of gases" or

4. on the "theory of stochastic processes",

5. intuitive techniques.

The Generalizations of Gibbs' ensemble formalism and Computational Modeling Methods are nowadays the most favorable approaches for providing satisfactory techniques for dealing with systems out of equilibrium.

The "Monte Carlo Method" and the "Nonequilibrium Molecular Dynamics" (NMD, for short) are part of what is know as *Computational Physics* or *Numeric Simulation Methods* [29-31]. The Nonequilibrium Molecular Dynamics is used for the study of properties of matter (in general, many-body systems) in which the direct integration of the dynamic equations of motion is done. B. J. Alder and T. E. Wainwright were the first to perform numerical simulation, done for a system of hard spheres [32]. Seven years later the case of molecules interacting through a Lenard-Jones potential was solved by A. Rahman [33]. Years later, R. Car and M. Parrinello improved the approach so-called as "*ab initio molecular dynamics*" [34].

Different kinetic theories were used to deal with the great variety of physical phenomena in nonequilibrium systems, in the absence of a Gibbs-style ensembles approach. It will be highlighted here the Non-Equilibrium Statistical Ensemble Formalism [10, 12, 35-37]. This formalism has an accompanying response function theory for nonequilibrium systems, a nonlinear quantum transport theory, a statistical thermodynamics plus a higher-order thermo-hydrodynamics. The Non-Equilibrium Statistical Ensemble Formalism (NESEF for short) is very powerful to deal with systems arbitrarily away from equilibrium. The present structure of the NESEF consists in a vast generalization and extension of earlier pioneering approaches. In this sense, the works of: J. G. Kirkwood [38], M. S. Green [39], H. Mori [35, 40], I. Oppenheim and J. Ross [35], and R. Zwanzig [41]. The NESEF was improved and systematized by the Russian School of statistical physics, highlighting: Nicolai Nicolaievich Bogoliubov [42], Nicolai Sergeievich Krylov [43], Dimitrii Zubarev [10], and Sergei Peletminskii [36, 37]; a systematization, generalizations and conceptual discussions of the NESEF are presented in Refs. [12] and [44].

These different approaches to the Non-Equilibrium Statistical Ensemble Formalism can be brought together under a unique variational principle. This procedure has been originally done by. D. N. Zubarev and V. P. Kalashnikov [45], and later on reconsidered in Ref. [12].

It should be emphasized that the NESEF can deal, within a certain degree of triumph, with some of the situations above described. The NESEF was applied, with a great success, in the following cases: non-linear thermo-hydrodynamics (including terms of highest order) in fluids under driven flow [46], in normal solutions and in complex situations as in solutions of microbatteries, DNA, micelles, and in polymers [47]; transport and optical processes in low-dimensional complex semiconductors [48, 49]; ultrafast optical spectroscopy [50]; non-linear transport in doped and in photoexcited polar semiconductors under electric fields [51, 52]; Bose-Einstein-Like Condensation [53].

Moreover, the generalized kinetic equations for far-from-equilibrium systems (with many-body) can be obtained by using the NESEF-based nonlinear kinetic theory [54]. In general, the kinetic equations for dynamical processes in many-body systems are dealt within certain approximations, for example, Vlasov equation for a plasma, Focker-Planck equation for a Brownian particle, Boltzmann equation for a dilute gas, the diffusion equation, Euler and Navier-Stokes equations for a compressible fluid, Landau equation for a weakly interacting gas, etc. [55]. Their common characteristic is that they all involve the dynamics of single-particle distributions. A question central of nonequilibrium statistical physics is the one of giving solid foundations to kinetic equations from microscopic dynamics and understanding their approximate validity. It should also be noted that one of the complicated problems of the theory of nonequilibrium transport processes in liquids and dense gases is the fact that their hydrodynamics and kinetics are coupled and must be treated simultaneously [56]. The microscopic descriptions of hydrodynamics, related to derivation of the kinetic equations from quantum or classical mechanics containing kinetic transport coefficients in terms of correlation functions, is a problem of long standing. The principal aspect is the derivation of constitutive laws which express thermodynamic fluxes, as those of energy and matter, in terms of thermodynamic forces. These laws (in their most general form) are non-instantaneous in time and non-local in space.

Moreover, a satisfactory construction of a "nonlinear kinetic theories" is very desirable for obtaining a deep physical insight on the physical phenomena governing these processes, which are fundamental to the development of nowadays modern technologies with economic and industrial relevance. These theories should allow to deal with: ultra-small systems, nanometric scale, low dimensional systems, nonlinear behavior, and ultrafast relaxation processes. In this case are involved the technologies for oil industry, cosmetics, food engineering, micelles, polymers, soft-matter engineering, electronic and opto-electronic devices, etc. It should be noted that the emergence of synergetic self-organization and instabilities may arise for systems in out of equilibrium conditions involving ultrafast relaxation processes, as in cases in laser-plasma interactions [57], semiconductor physics [58] and biophysics [59]. The formalism can be extended to deal with anomalous situations which are associated to disordered media, for example: systems showing a fractal-like characteristics, distinctive behavior of polyatomic structures such as surfactant micelles, polymers, colloidal particles, DNA, biopolymers in liquid solutions [15-17].

## 3. Conclusions

Many situations involving manufacturing processes and technological applications, in general, are associated to nonlinear transport and optical properties in systems far from equilibrium, being in constrained geometries and presenting ultrafast relaxation processes and nonlinear behavior.

Here the fundamental point in the scientific method, of corroborating theory by comparison with experience, can not be forgotten [27]. S. J. Gould's pointed out that: "a detail, by itself, is blind; a concept without a concrete illustration is empty", while, Charles Robert Darwin expressed that: "theory and observation are siamese twins, inextricably intertwined and continually interacting" [60]. On this question, R. Kubo pointed out that:

> "... statistical mechanics exists for the real world, not for fictions. Further progress only can be achieved with close cooperation with experience" [28]. The English theoretical physicist, cosmologist, Stephen Hawking expressed that: "I do not demand that a theory corresponds to reality. [...]. I do not demand that a theory corresponds to reality [...]. All I am concerned with is that the theory should predict the results of measurement" [61].

It should be reinforced that: boldness and imagination in making theory must always be policed by experience. Any theory in Physics, for its validation, requires to show a good agreement between "calculated values" (theory) and "measured values" (experimental data) [62, 63].

Closing this paper, seven areas are mentioned. These areas are of particular interest, at the scientific, technological and at the production line, and therefore of relevance for government and society, successfully analyzed in terms of the formalism just describe. Summarizing, they are:

1. nonconventional thermo-hydrodynamics;

2. ultrafast relaxation processes in semiconductors devices;

3. nonequilibrium Bose-Einstein-like condensations and coherent states;

4. low-dimensional semiconductors;

5. thermo-statistics of complex structured systems;

6. nonlinear higher-order thermo-hydrodynamics; and

7. nonlinear transport in semiconductors devices.